\documentclass[aps,prb,twocolumn,superscriptaddress]{revtex4-1}
\usepackage{graphicx}
\usepackage[colorlinks,linkcolor=blue,citecolor=blue,urlcolor=blue,hyperindex,pdfstartview=FitH,plainpages=false]{hyperref}

\newcommand {\uJcm}{$\mu$J/cm$^2$}

\begin{document}
\bibliographystyle{unsrt}


\title{Optical manipulation of the topological phase in ZrTe$_5$ revealed by time- and angle-resolved photoemission}
\author{Chaozhi Huang}
\email{These authors contributed equally.}
\author{Chengyang Xu}
\email{These authors contributed equally.}
\author{Fengfeng Zhu}
\author{Shaofeng Duan}
\author{Jianzhe Liu}
\author{Lingxiao Gu}
\author{Shichong Wang}
\author{Haoran Liu}
\affiliation{Key Laboratory of Artificial Structures and Quantum Control (Ministry of Education), School of Physics and Astronomy, Shanghai Jiao Tong University, Shanghai 200240, China}
\author{Dong Qian}
\email{dqian@sjtu.edu.cn}
\affiliation{Key Laboratory of Artificial Structures and Quantum Control (Ministry of Education), School of Physics and Astronomy, Shanghai Jiao Tong University, Shanghai 200240, China}
\affiliation{Collaborative Innovation Center of Advanced Microstructures, Nanjing University, Nanjing 210093, China}
\affiliation{Tsung-Dao Lee Institute, Shanghai Jiao Tong University, Shanghai 200240, China}
\author{Weidong Luo}
\email{wdluo@sjtu.edu.cn}
\author{Wentao Zhang}
\email{wentaozhang@sjtu.edu.cn}
\affiliation{Key Laboratory of Artificial Structures and Quantum Control (Ministry of Education), School of Physics and Astronomy, Shanghai Jiao Tong University, Shanghai 200240, China}
\affiliation{Collaborative Innovation Center of Advanced Microstructures, Nanjing University, Nanjing 210093, China}

\date {\today}

\begin{abstract}
High-resolution time- and angle-resolved photoemission measurements were conducted on the topological insulator ZrTe$_5$. With strong femtosecond photoexcitation, a possible ultrafast phase transition from a weak to a strong topological insulating phase was experimentally realized by recovering the energy gap inversion in a time scale that was shorter than 0.15 ps. This photoinduced transient strong topological phase can last longer than 2 ps at the highest excitation fluence studied, and it cannot be attributed to the photoinduced heating of electrons or modification of the conduction band filling. Additionally, the measured unoccupied electronic states are consistent with the first-principles calculation based on experimental crystal lattice constants, which favor a strong topological insulating phase. These findings provide new insights into the longstanding controversy about the strong and weak topological properties in ZrTe$_5$, and they suggest that many-body effects including electron--electron interactions must be taken into account to understand the equilibrium weak topological insulating phase in ZrTe$_5$.


\end{abstract}


\maketitle

\section{Introduction}

Significant research has been devoted to investigating topological materials and understanding the transitions among different topological phases, since it can broaden the understanding of new states of matter and have potential applications in electronic and spintronic devices.
Transitions between topologically non-trivial and trivial states can be induced by applying external strain \cite{Bruene2011,Lin2021}, magnetic fields \cite{Zheng2019,Cao2015}, pressure \cite{Xi2013,Ideue2019}, thermal lattice expansion \cite{Dziawa2012}, and chemical doping \cite{Xu2011,Sato2011,Brahlek2012,Xu2015a}.
The optical manipulation of topological phase transitions has recently garnered significant attention because of its scientific and practical significance. Various theoretical studies have explored the possibility of optically driven topological phase transitions in different systems \cite{Inoue2010,Inoue2012,Ezawa2013,Chen2016,Ezawa2017,Ledwith2018,Liu2018,Tanaka2021,Vaswani2020,Shao2021}. However, experimental realization of the optically driven topological phase transitions is quite challenging because of the lack of suitable platforms.

The quasi-one-dimensional material ZrTe$_5$ has a layered orthorhombic crystal structure with a space group of $Cmcm$ (Fig.~\ref{Fig1}(a)). This material gained attention due to an anomalous resistance peak at a specific temperature, which was accompanied by a sign change in thermopower \cite{Okada1980,Littleton1998}. It was predicted to be a three-dimensional topological insulator (TI) near the phase boundary between a strong topological insulator (STI) with gapped bulk bands and surface states on every surface, and a weak topological insulator (WTI) with surface states only on side surfaces.\cite{Weng2014}. Subsequent experimental work has shown considerable discrepancies regarding the topological properties of the electronic states, which can exhibit the behavior of STI \cite{Manzoni2016,Jiang2020}, WTI \cite{Li2016a,Wu2016,Zhang2017,Xiong2017}, or a Dirac semimetal \cite{Chen2015a,Yuan2016,Li2016,Zheng2016,Shen2017}. Recently, the identification of the three-dimensional quantum Hall effect confirmed the presence of nontrivial topological electronic states in ZrTe$_5$ \cite{Tang2019}. However, the WTI behavior was inconsistent with calculations based on the experimental lattice structure but consistent with the slightly expanded (calculation optimized) lattice constants instead \cite{Weng2014,Wu2016}. These discrepancies may result from various experimental conditions, including the measuring temperature, sample strain, and dopings, which suggests the potential tunability of the topological phase in ZrTe$_5$. The topological phase transitions in ZrTe$_5$ can be realized by tuning the temperature \cite{Xu2018} and applying an external magnetic field \cite{Zheng2017}, pressure \cite{Zhou2016}, and strain \cite{Zhang2021,Mutch2023}. Along with these equilibrium methods, phase transitions at ultrafast time scales can be induced by ultrafast laser excitation through light--matter interactions. Such topological phase transition triggered by the photoinduced coherent phonon modes or atomic motion in ZrTe$_5$ has been experimentally and theoretically proposed \cite{Vaswani2020,Konstantinova2020,Aryal2021}, but the direct measurement of the electronic structures during this intriguing phenomenon remains lacking. Searching for electronic evidence of the photoinduced topological phase transition and establishing the underlying ultrafast phase transition physics are important to the field.

In this Letter, we present signatures of an ultrafast photoinduced phase transition in ZrTe$_5$ from a weak to a strong topological insulating phase investigated by time- and angle-resolved photoemission spectroscopy (TRARPES).
In particular, this topological phase transition was characterized by a photoinduced energy gap inversion, and the non-equilibrium STI phase could last for over 2 ps at the highest pump fluence studied.
Interestingly, the precisely characterized unoccupied states were consistent with the bulk bands of the STI phase from first-principles calculations based on the experimental crystal lattice constants, although the experimental equilibrium states favor a weak insulating behavior. The observed manipulation of the topological phase cannot be explained by the photoinduced heating of electrons or modification of the conduction band filling from first-principles calculations. Rather, the screening of Coulomb interactions by photoinduced charge carriers may drive the phase transition, which indicates that many-body effects, including strong electron--electron interactions, may play crucial roles in establishing the weak topological insulating phase in ZrTe$_5$.

\section{Methods}

The TRARPES experiments were performed on a home-built TRARPES system \cite{Yang2019,Huang2022} using a 1.77-eV infrared laser as the pump beam and a 6.05-eV ultraviolet laser as the probe beam. The measurements were conducted with a repetition rate of 250 kHz, and the spot size on the sample for the pump and probe beams was approximately 90 and 23 $\mu$m, respectively. The best overall time resolution and energy resolution of the setup were approximately 113 fs and 16 meV, respectively. It is noteworthy that a high probe photon flux was used in the time-delay measurements to reach reasonable photoemission count rates and the estimated energy resolution was about 60 meV due to the presence of the space charge effect. High-quality single crystals of ZrTe$_5$ were grown using the chemical vapor transport method. The sample was cleaved at 25 K in an ultrahigh vacuum with a base pressure better than 3$\times$10$^{-11}$ Torr. The energy shifts induced by the space charge and surface photovoltage effects during the photoemission process were corrected as discussed in the Supplemental Material, Discussion No. 1.

\section{Results}
\subsection{Unoccupied states}

\begin{figure}[b]
\centering
\includegraphics[width=8cm]{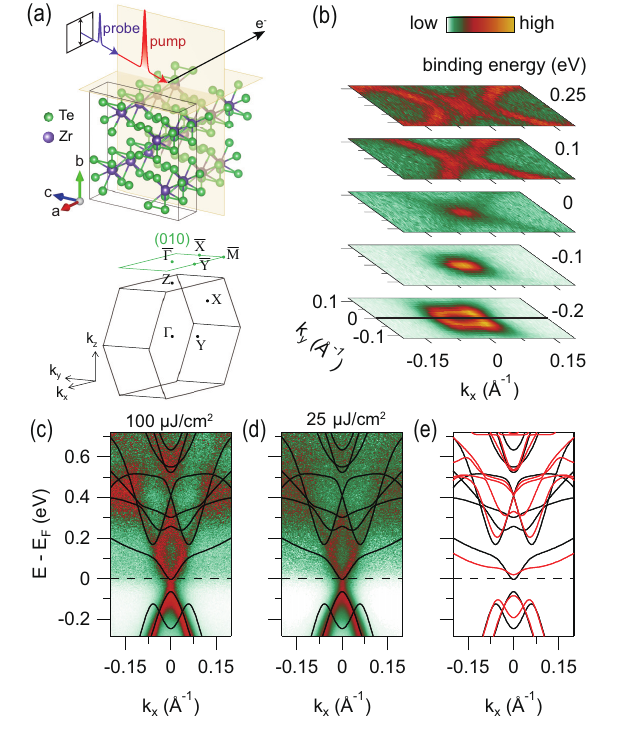}
\caption{
(a) A schematic of the TRARPES experimental set-up, crystal structure, bulk Brillouin zone, and projected (010) surface Brillouin zone of ZrTe$_5$. The polarization of the pump and probe beams are both perpendicular to the a axis of the crystal. 
(b) Constant-energy mappings at different energy cuts from -0.2 to 0.25 eV. 
(c) Band structure measured along $\bar{\Gamma}$ - $\bar{\rm X}$ ($k_x$) as indicated by the black line in (b) and the calculated bulk bands based on the experimental lattice constant plotted on top of the spectra with an offset of -0.04 eV. The spectra in (b) and (c) were measured at a delay time near time 0 with a pump fluence of 100 \uJcm. The intensities were normalized to the height of the momentum distribution curve to better illustrate the unoccupied states.
(d) Same as (c) but measured with a moderate pump fluence of 25 $\mu$J/cm$^2$.
(e) Calculated bulk bands for the STI (black curves, experimental lattice constant) and WTI (red curves, optimized lattice constant) with an offset of -0.04 eV.
}
\label{Fig1}
\end{figure}

With improved time and energy resolutions, we could clearly identify the occupied and unoccupied electronic states of ZrTe$_5$ around the $\Gamma$ point near the time of 0 with a pump fluence of 100 \uJcm. The constant-energy maps taken at binding energies from -0.2 to 0 eV were consistent with previous reports, which show that the curved rectangle contour that corresponds to the valence band (VB) gradually shrinks to a small point \cite{Xiong2017,Zhang2017} (Fig.\ref{Fig1}(b)).
The measured band structure above the Fermi energy along $\bar{\Gamma}$ - $\bar{\rm X}$ ($k_x$) from approximately -0.3 to 0.8 eV matched the first-principles calculation based on the experimental lattice constants (Fig.~\ref{Fig1}(c), Supplemental Materials, Discussion No. 2). By contrast, the calculated band structure based on the optimized lattice constants clearly deviated from the experimental unoccupied state for the conduction band near the Fermi energy and the band near the momentum $\pm$ 0.1 \AA$^{-1}$ and 0.4 eV above the Fermi energy (Figs.~\ref{Fig1}(c) and (e)). The band structure measured with a moderate pump fluence of 25 $\mu$J/cm$^2$ exhibits similar characteristics, generally aligning well with a STI (Fig.~\ref{Fig1}(d)). We note that the calculations were performed based on density functional theory (DFT) with the spin-orbit-coupling effect and $k_z$ effect. The measurement of the unoccupied band structure indicates that the experimental lattice configuration of ZrTe$_5$ favors a STI, but many recent experiments have suggested that it is a WTI instead \cite{Wu2016,Zhang2017,Xiong2017,Zhang2021}. To resolve this controversy, we performed ultrafast dynamics measurements on the electronic states near the Fermi energy in ZrTe$_5$, and we elaborate on this below.

\subsection{Temporal evolution}

\begin{figure}[b]
\centering
\includegraphics[width=8cm]{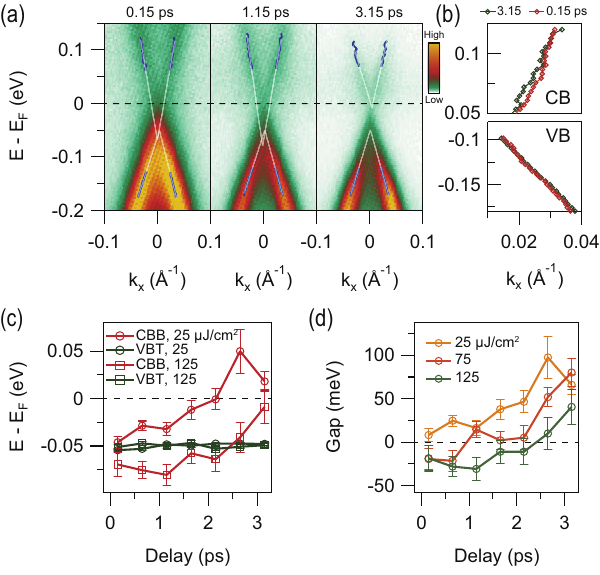}
\caption{
(a) Near-Fermi-energy electronic structures measured with the pump fluence of 125 \uJcm~at 0.15, 1.15, and 3.15 ps. The thick blue lines denote the band dispersions extracted from peak positions of the momentum dispersion curves. The white lines are the fittings to the linear functions. The fitted momentum of the CBB and VBT are offset and not at the zero momentum possibly due to slightly misaligned sample orientation. To improve the statistics of the data above the Fermi energy, the spectra were averaged within a time window of 0.4 ps when extracting MDCs of the conduction band (CB).
(b) The average MDC dispersions of the CB and VB at 0.15 and 3.15 ps.
(c) Energy position of the CBB and VBT as a function of the delay time with the pump fluence of 25 and 125 \uJcm, which was estimated by linearly extrapolating the high-energy dispersion. 
(d) Nominal gap size from the energy difference between CBB and VBT at different pump fluences as a function of the delay time.
}
\label{Fig2}
\end{figure}

After the pumping with an intense fluence of 125 \uJcm, the CB and VB near the Fermi energy moved toward each other and a nominal negative energy gap of approximately -20 meV appeared at 0.15 ps and such negative energy gap lasted longer than 2 ps after photoexcitation and the nominal gap became positive at a longer delay time of 3.15 ps (Figs. \ref{Fig2}(a) and (d)). 
The excited CB bottom remained resolvable up to 3.15 ps, and the extracted energy gap was approximately 40 meV, which is qualitatively consistent with previous equilibrium experiments \cite{Zhang2017,Zhang2019,Zhang2021}. 
After the pumping with a moderate fluence of 25 \uJcm, the CB and VB maintained a significant gap of approximately 10 meV at 0.15 ps, and they gradually separated at longer delay times (Fig. \ref{Fig2}(c)). 
The observed shrinkage of the energy gap after photoexcitation can be explained by the combined effects of the increased electron velocities and shifts of the CB and VB bands (Fig. \ref{Fig2}(b)). All the CB bottom (CBB) and VB top (VBT) were extracted by linearly extrapolating the dispersion derived from the momentum distribution curves (MDCs) between the same momentum $\pm$0.032 and $\pm$0.022 \AA$^{-1}$ for the CB and between momentum about $\pm$0.036 and $\pm$0.02 \AA$^{-1}$ for the VB, respectively. This method for estimating the energy gap was widely used in previous reports \cite{Xiong2017,Zhang2021}. 
We note that the fitted momentum of the CBB and VBT are offset and not at zero momentum. This is probably due to the fact that the constant-energy contour is not symmetric in all directions, and there was a slight misalignment in the tilt angle and in-plane angle, accompanied by a fitting error.

According to the observed indicators of photoinduced inversion of the energy gap, it is reasonable to conclude that strong photoexcitation potentially drives an ultrafast phase transition from the weak topological insulating phase to the underlying strong insulating phase in ZrTe$_5$. This conclusion is supported by several factors: 1) previous experiments have shown evidence of an equilibrium weak topological insulating phase \cite{Li2016a,Wu2016,Zhang2017,Xiong2017}; 2) calculations based on experimental lattice constants suggest a strong insulating behavior, since the calculated electronic structure very well matches the experimental unoccupied states (Fig.~\ref{Fig1}(c)); 3) a negative bulk energy gap in a strong insulating phase has been predicted by various previous theoretical works \cite{Weng2014,Vaswani2020,Zhang2021,Aryal2021,Mutch2023}. 

\subsection{Fluence dependence}

\begin{figure}[t]
\centering
\includegraphics[width=8.6cm]{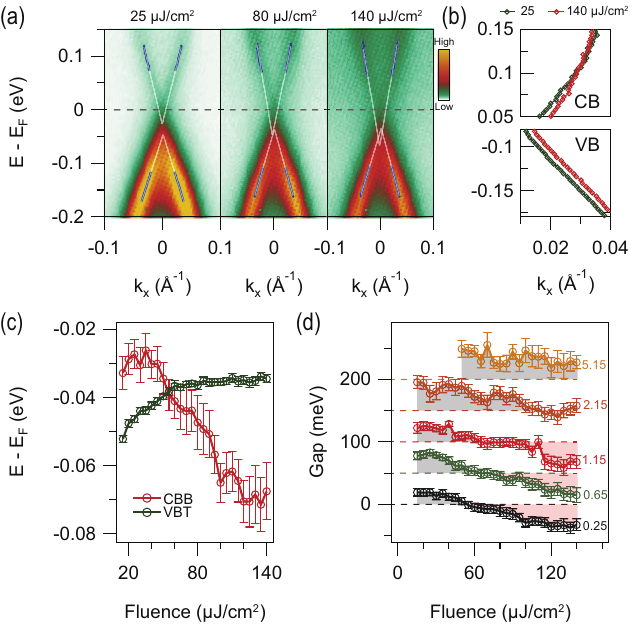}
\caption{
(a) Non-equilibrium electronic structures of ZrTe$_5$ measured along $k_x$ at 0.25 ps with pump fluences 25, 80, and 140 \uJcm. 
(b) The average MDC dispersions of the CB and VB with pump fluence 25 and 140 \uJcm~at 0.25 ps.
(c) Energies of the CBB and VBT as functions of the pump fluence at 0.25 ps. 
(d) The fluence-dependent bulk gap sizes estimated by linearly extrapolating the MDC dispersions measured at different delay times as noted by the numbers. Each curve is offset by 50 meV vertically. The CB cannot be resolved clearly below 50 \uJcm~at 5.15 ps. The gray and red shaded areas indicate the positive and negative gap, respectively.
}
\label{Fig3}
\end{figure}

The critical pump fluence for the observed gap inversion behavior was determined by conducting fluence-dependent measurements of the energy gap between CB and VB with better data quality. At a delay time of 0.25 ps, a positive energy gap was observed for the pump fluence of 25 \uJcm, while the gap was closed or even negative for higher pump fluences of 80 and 140 \uJcm~(Fig. \ref{Fig3}(a)). Considering the energy gap size might be underestimated by linearly extrapolating the dispersions, we apply a more complex model to fit the MDC dispersions (Supplemental Materials, Discussion No. 3). The results also yield a fully closed gap at a high pump fluence immediately following photoexcitation. The pump-induced large downshift of the CB can be also indicated by the enhanced absolute Fermi velocity (6.8$\times$10$^{5}$ and 8.8$\times$10$^{5}$ m/s at the fluence of 25 and 140 \uJcm, respectively) (Fig. \ref{Fig3}(b)). The energies of the CBB and VBT as functions of the pump fluence clearly indicate an energy gap inversion around 60 \uJcm~at 0.25 ps (Fig. \ref{Fig3}(c)). The critical pump fluence was also observed in the extracted energy gap as a function of the pump fluence for the delay times of 0.65, 1.15, and 2.15 ps but absent for delay times at 5.15 ps (Fig.~\ref{Fig3}(d)). Although the fluence resolution was reduced during the experiments to obtain reasonable statistics for the MDC dispersion analysis in the unoccupied states, the critical pump fluence remained higher at longer delay times, which suggests that the observed energy gap is potentially related to the photoinduced non-equilibrium charge carriers, which should exponentially decay as a function of the delay time. The photoinduced shrinkage of the energy gap saturates above about 120 \uJcm~ at 0.25 ps (Figs. \ref{Fig3}(c) and (d)), giving a fully inverted gap of about 35 meV.

\section{Disccusion}

\begin{figure}
\centering\includegraphics[width=0.9\columnwidth]{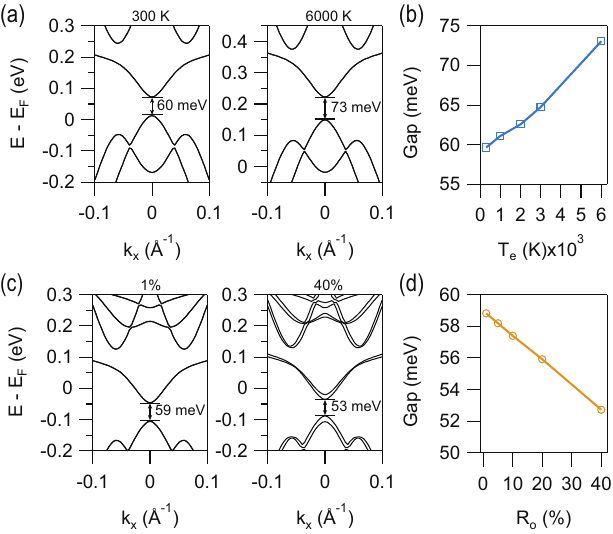}
\caption{
(a) Calculated band structures in the STI case for electronic temperature T$_e$ at 300 and 6000 K.
(b) Gap size at $\Gamma$ as a function of T$_e$. 
(c) Calculated band structures in the STI case for different occupation rates (R$_o$) in CBs at 1$\%$ and 40$\%$ at 0 K. The band splitting for R$_o$ = 40\% results from the enhanced spin-orbit coupling by increasing R$_o$. 
(d) Gap size as a function of R$_o$.
}
\label{Fig4}
\end{figure}

Several possible mechanisms for the experimentally observed optical manipulation of the topological phase can be excluded. During the photoexcitation process, the electronic temperature (T$_e$) can be temporarily heated to thousands of degrees due to electron--electron scattering for the pump photon energy and fluence in the experiments. However, the effect of T$_e$ on the phase transition can be excluded because the bulk gap size only slightly increased from 60 to 73 meV when the electronic temperature was increased from 300 to 6000 K, as shown in the calculated near-Fermi-energy band structures in Figs. \ref{Fig4}(a) and (b). Additional calculations from very low temperature to 6000 K demonstrate that the electronic-temperature-driven energy gap monotonically increases when T$_e$ increases (Fig. \ref{Fig4}(b)), which contradicts these experimental observations. Then, we altered the occupancy of the VB and CB as an analogy of the excited state of the system to calculate the bulk energy gap as a function of the occupation rate (R$_o$) (Figs. \ref{Fig4}(c) and (d)). However, the gap size is only reduced by a few meV up to R$_o$ of 40\%. For the pump fluences in the experiments, the estimated photoinduced change in occupancy was less than 2\%, which is far from driving the tens-meV change in the energy gap observed in the experiments. In addition, specific coherent phonon modes can induce a phase transition between STI and WTI states \cite{Vaswani2020,Konstantinova2020,Aryal2021}, but the magnitude of the energy shift induced by the coherent phonon mode of 1.2 THz is less than 3 meV (Supplemental Materials, FIG. 5), which is much smaller than the photoinduced energy gap change in Figs.~\ref{Fig2} and ~\ref{Fig3}. Moreover, the observed gap inversion behavior is unlikely to be attributed to the surface photovoltage effect, given its usual duration of several tens of picoseconds\cite{Kremer2021}. Finally, it cannot be a result of atomic motions in the crystal lattice, since the atomic position is almost unchanged in this short timescale of picoseconds \cite{Konstantinova2020}.

The existence of non-negligible Coulomb interactions may play a crucial role in establishing the WTI phase, since in a Dirac system, spontaneous electron--hole pair formation through Coulomb interactions is favored due to the Dirac electron--hole symmetry \cite{Kotov2012,Janssen2016,Wang2017}. Indeed, recent nuclear magnetic resonance studies on ZrTe$_5$ have shown evidence of exciton formation near the Fermi energy \cite{Tian2021}. The previously observed WTI characteristics in equilibrium studies on ZrTe$_5$ may occur because a sizable Coulomb interaction breaks the energy gap inversion, although the experimental lattice constants favor an STI based on the first-principle calculation. In this context, we propose that upon photoexcitation, the photoinduced non-equilibrium charge carriers strengthen the screening of Coulomb interactions, as observed in TiSe$_2$ \cite{Rohwer2011,Mathias2016}, GaAs \cite{Huber2001}, and Ta$_2$NiSe$_5$ \cite{Tang2020}, and the inverted energy gap and associated STI can be restored above the critical pump fluence, where photoinduced non-equilibrium charge carriers can sufficiently screen the possible electron--hole interaction. However, additional theoretical studies with sizable Coulomb interaction or other many-body effects considered are necessary to confirm this picture.

In conclusion, we have investigated the ultrafast electronic dynamics of ZrTe$_5$ and provided evidence of an optical manipulation of its topological phase. Such an ultrafast phase transition is not only physically interesting but also has potential novel applications in the future high-speed electronic devices. The photoexcited charge carrier screening appears to be a plausible mechanism for this phase manipulation. Moreover, our findings provide new insights into the longstanding controversy regarding the strong and weak topological properties in ZrTe$_5$. Many-body effects including electron--electron interactions may have driven ZrTe$_5$ out of the strong topological insulating phase that the first-principles calculations based on experimental lattice constants predict, which results in a weak topological insulating phase.

\section{Acknowledgements}

W. T. Z. acknowledges support from the National Key R\&D Program of China (Grants No. 2021YFA1400202 and No. 2021YFA1401800) and National Natural Science Foundation of China (Grants No. 12141404 and No. 11974243) and Natural Science Foundation of Shanghai (Grants No. 22ZR1479700 and No. 23XD1422200). S. F. D. acknowledges support from the China Postdoctoral Science Foundation (Grant No. 2022M722108). D.Q. acknowledges support from the National Key R\&D Program of China (Grants No. 2022YFA1402400 and No. 2021YFA1400100) and National Natural Science Foundation of China (Grant No. 12074248).

\end{document}